\documentclass[twocolumn,aps,prb,superscriptaddress]{revtex4}

\usepackage{graphicx}
\usepackage{dcolumn}
\usepackage{bm}
\usepackage[cmex10]{amsmath}
\usepackage{sidecap}

\usepackage[utf8]{inputenc}
\usepackage[T1]{fontenc}

\begin{document}

\preprint{APS/123-QED}

\title{Temperature dependence of spin-orbit torques in W/CoFeB bilayers}

\author{Witold Skowro\'{n}ski}
 \email{skowron@agh.edu.pl}
\author{Monika Cecot}
\author{Jarosław Kanak}
\author{Sławomir Ziętek}
\author{Tomasz Stobiecki}
\affiliation{AGH University of Science and Technology, Department of Electronics, Al. Mickiewicza 30, 30-059 Krak\'{o}w, Poland}
\author{Lide Yao}
\author{Sebastiaan van Dijken}
\affiliation{NanoSpin, Department of Applied Physics, Aalto University School of Science, P.O.Box 15100, FI-00076 Aalto, Finland}
\author{Takayuki Nozaki}
\author{Kay Yakushiji}
\author{Shinji Yuasa}
\affiliation{National Institute of Advanced Industrial Science
and Technology, Spintronics Research Center, Tsukuba, Ibaraki 305-8568, Japan}

\date{\today}

\begin{abstract}
We report on the temperature and layer thickness variation of spin-orbit torques in perpendicularly magnetized W/CoFeB bilayers. Harmonic Hall voltage measurements reveal dissimilar temperature evolutions of longitudinal and transverse effective magnetic field components. The transverse effective field changes sign at 250
K for a 2 nm thick W buffer layer, indicating a much stronger contribution from interface spin-orbit interactions compared to, for example, Ta. Transmission electron microscopy measurements reveal that considerable interface mixing between W and CoFeB is primarily responsible for this effect.
\end{abstract}

\maketitle

Efficient manipulation of magnetization using electrical signals at the nanoscale will further the development of next generation magnetic memories, logic \cite{kawahara_spin-transfer_2012} and microwave devices \cite{stamps_2014_2014}. 
The spin Hall effect\cite{sinova_universal_2004} and Rashba effect\cite{miron_current-driven_2010} are intensively studied, as they produce effective magnetic fields that can be used to switch the magnetization of magnetic nano-pillars \cite{miron_perpendicular_2011}, excite microwave oscillations in nano-discs \cite{demidov_magnetic_2012} or induce magnetic domain-wall motion in nanowires \cite{emori_current-driven_2013}. 
Quantitatively, spin-orbit torques are characterized by the spin Hall angle ($\theta_\mathrm{H}$), which is a measure of the ratio between spin current density ($J_\mathrm{s}$) and charge current density ($J_\mathrm{c}$). Different heavy metal/ferromagnet bilayers have been proposed as the source of spin-orbit torque, including: Ta \cite{kim_layer_2012}, Hf \cite{akyol_effect_2015}, Pt \cite{liu_spin-torque_2011}, CuIr \cite{niimi_extrinsic_2011}, and W \cite{pai_enhancement_2014, cho_large_2015}, with W exhibiting the largest $\theta_\mathrm{H}$ to date. Recent work on oxidized W has also revealed promising results \cite{demasius_enhanced_2016}. 
Studies on symmetry of spin-orbit torques pointed out different contributions to the effective magnetic field arising from the bulk spin Hall effect and interface Rashba interactions \cite{garello_symmetry_2013}. In addition, it has been shown that the interface between heavy metal and ferromagnetic layers strongly affects both $\theta_\mathrm{H}$ \cite{zhang_role_2015} and the spin diffusion length \cite{Rojas_Sanchez_spin_2014}. 


In this letter, we demonstrate that spin-orbit torques in W/CoFeB bilayers are strongly affected by interface mixing. Both transverse and longitudinal torque components are measured in a temperature range from 19 to 300 K. The interface contribution to the transverse effective field is dominant at low temperatures, leading to a sign reversal at 250 K, where the bulk spin-orbit torque starts to dominate. Considerable mixing between W and CoFeB is supported by transmission electron microscopy (TEM) and X-ray reflectivity (XRR) analysis. 

The investigated samples consisted of sputter-deposited multilayers with the following structure: W($t_\mathrm{W}$)/Co$_\mathrm{12}$Fe$_\mathrm{68}$B$_\mathrm{20}$(1.3)/MgO(2.5)/Ta(4) (thicknesses in nm), with $t_\mathrm{W}$ = 2, 4, and 6 nm. After deposition, the samples were measured using vibrating sample magnetometry (VSM), X-ray diffraction (XRD) and XRR and they were successively annealed in a high vacuum chamber. Microstructure analysis was performed using a JEOL 2200FS TEM with double Cs correctors, operated at 200 keV. Cross-sectional TEM specimens were prepared by a MultiPrep polishing machine (Allied High-Tech) and Ar ion milling. Selected samples were patterned using e-beam lithography, ion-beam etching and lift-off processes into 70 $\mu$m long Hall bars of different width spanning from 1 to 40 $\mu$m. During microfabrication, electrical contacts with a dimensions of 100 $\times$ 100 $\mu$m were deposited and most of the Ta top layer was etched away leaving only a thin oxidized layer as protection.

The resistivity of the samples was measured using a four-probe method, both for as-deposited devices (with in-plane magnetic anisotropy) and annealed ones (with effective perpendicular magnetic anisotropy - PMA). 
The harmonic Hall voltage measurements were carried out using lock-in amplification in a Janis cryogenic probe station equipped with an electromagnet. During these experiments, the temperature was varied between 19 and 300 K. Measurements were performed for various magnetic field orientations: perpendicular to the sample plane (along $z$ axis - i.e. anomalous Hall effect (AHE) configuration), longitudinal to the stripe (along $x$ axis) and transverse to the stripe (along $y$ axis).

\begin{figure}
\centering
\includegraphics[width=\columnwidth]{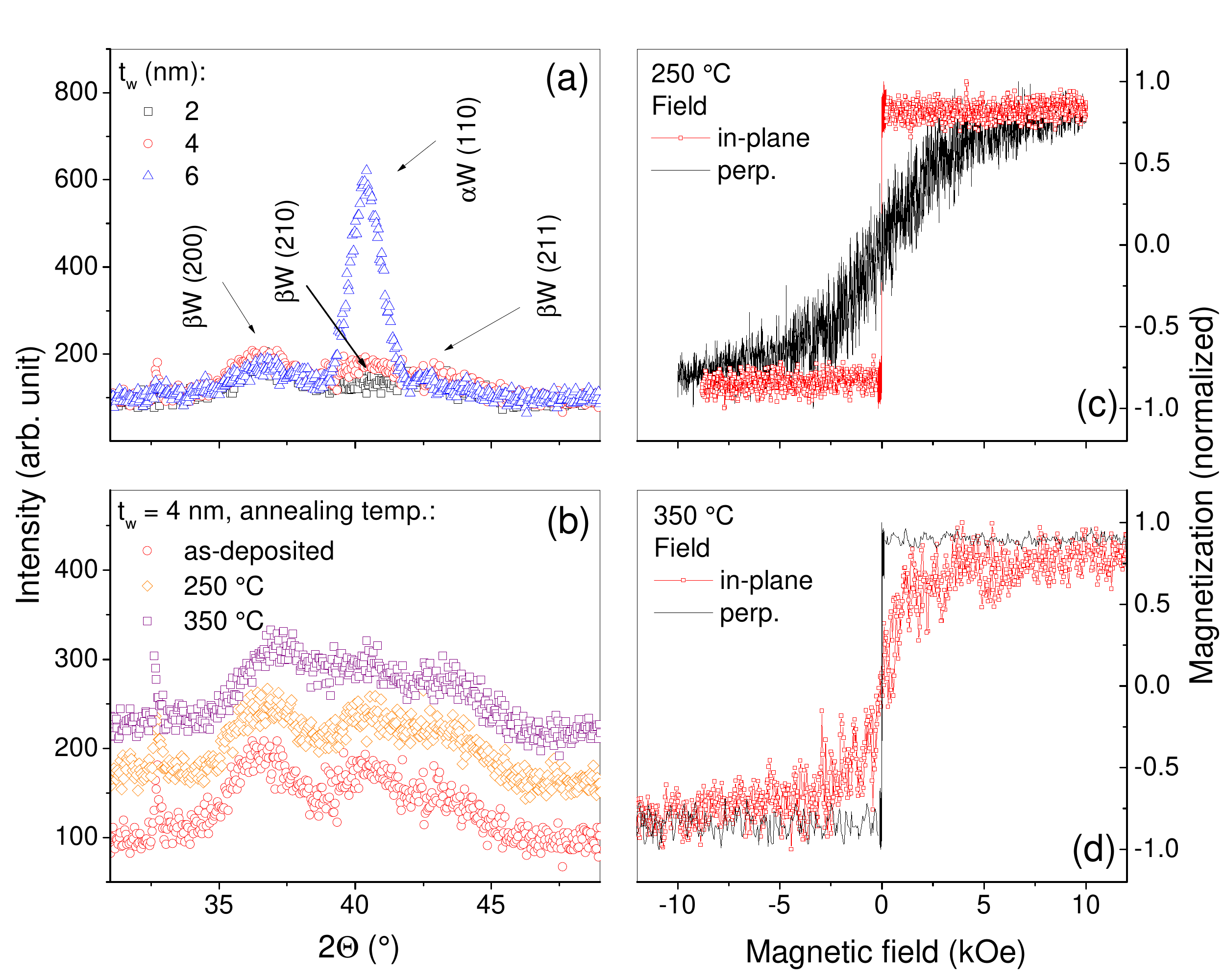}
\caption{XRD $\theta$-2$\theta$ scans for as-deposited samples with different $t_\mathrm{W}$ (a) and for annealed samples with $t_\mathrm{W}$ = 4 nm (b) – curves are offset for clarity. In-plane and perpendicular magnetization curves for the sample with $t_\mathrm{W}$ = 4 nm after annealing at 250 $^\circ$C (c) and 350 $^\circ$C (d).}
\label{fig:xrd_vsm}
\end{figure}

First, the crystallographic phases of W with different layer thicknesses ($t_\mathrm{W}$ = 2, 4, and 6 nm) were determined using XRD measurements. The $\theta$-2$\theta$ scans of Fig. \ref{fig:xrd_vsm}(a) reveal that the $\beta$-tungsten phase is present in all samples, whereas, a clear $\alpha$-tungsten reflection is visible only for the sample with a 6 nm thick W layer. 
The same conclusion can be drawn from four-point resistivity measurements. Assuming a CoFeB resistivity of 113 $\mu\Omega\cdot$cm (measured independently) and a parallel resistor model, the calculated W resistivity ($\rho_\mathrm{W}$) amounts 128 and 105 $\mu\Omega\cdot$cm for $t_\mathrm{W}$ = 2 nm and 4 nm, respectively, but it decreases significantly to 36 $\mu\Omega\cdot$cm for $t_\mathrm{W}$ = 6 nm, supporting the presence of a low-resistive $\alpha$-tungsten phase \cite{pai_spin_2012}. We also verified, that $\alpha$-tungsten does not form in the thinner W layers during annealing (Fig. \ref{fig:xrd_vsm}(b)). 
Because the $\beta$-tungsten phase is crucial for obtaining PMA and large spin Hall angles \cite{hao_giant_2015}, we focus on the thinner W buffers only. We note that the resistivity of W/CoFeB bilayers did not change (within experimental error) upon annealing.

\begin{figure}
\centering
\includegraphics[width=\columnwidth]{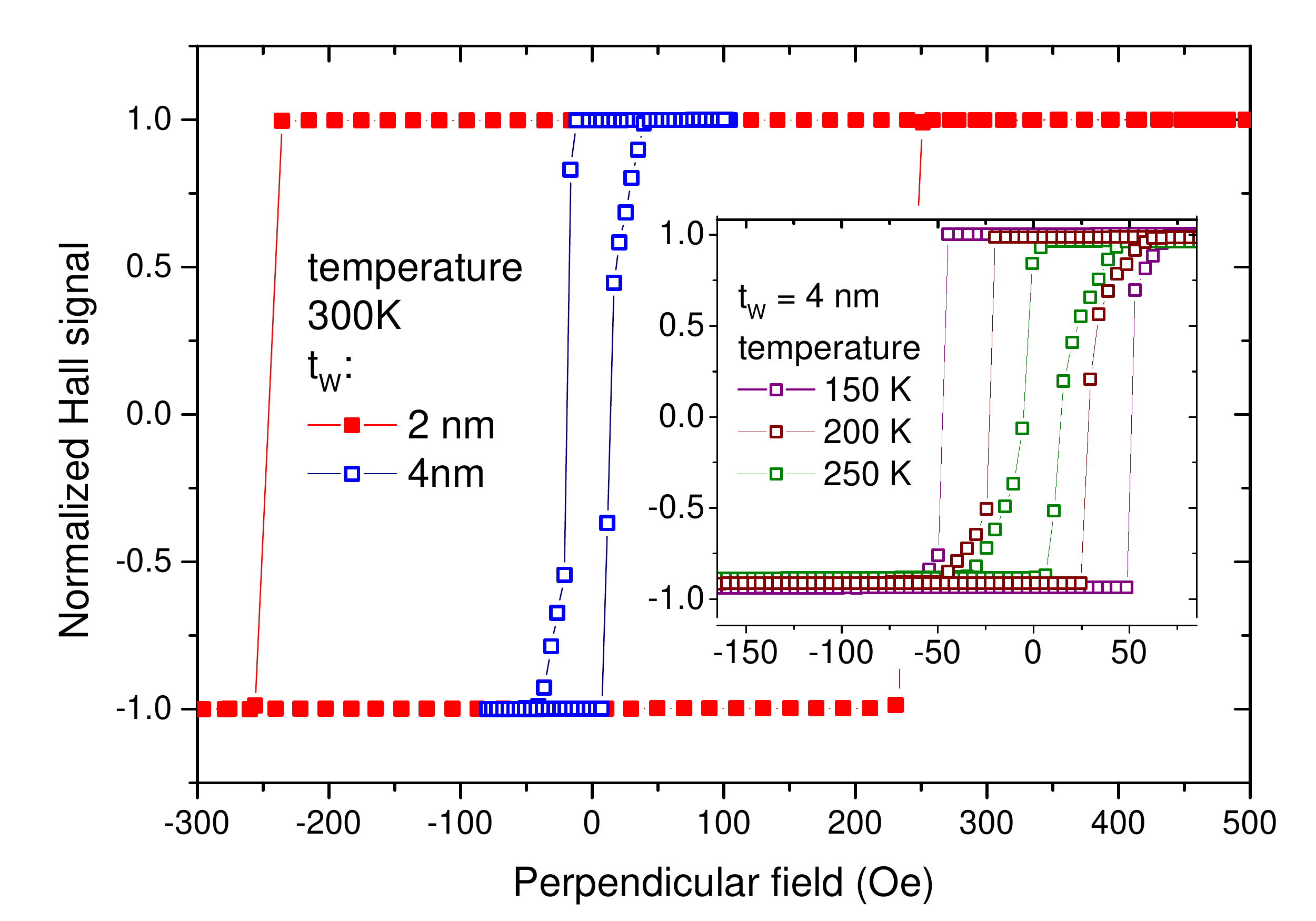}
\caption{AHE vs. perpendicular magnetic field measured in a room temperature for samples with $t_\mathrm{W}$ = 2 (full symbols) and 4 nm (open symbols). Inset presents AHE measured for $t_\mathrm{W}$ = 4 nm measured at various temperatures.}
\label{fig:ahe}
\end{figure}

Independently, the magnetic properties of the deposited stacks were verified using VSM. Exemplary magnetic hysteresis loops for $t_\mathrm{W}$ = 4 nm are presented in Fig. \ref{fig:xrd_vsm}. The measurements indicate a transition from in-plane anisotropy to PMA after annealing at 350 $^\circ$C, which is consistent with our earlier work \cite{skowronski_underlayer_2015}.

Next, the AHE of samples with $t_\mathrm{W}$ = 2 and 4 nm were measured in perpendicular magnetic field (Fig. \ref{fig:ahe}). Although both W layers induce PMA in CoFeB, the switching for $t_\mathrm{W}$ = 2 nm is more abrupt compared to $t_\mathrm{W}$ = 4 nm. This effect, which can be attributed to more gradual magnetization rotation or magnetic domain formations\cite{torrejon_interface_2014}, hampers the extraction of effective magnetic fields. We note that this behavior persist even at low temperatures (inset in Fig. \ref{fig:ahe}), ruling out superparamagnetism in 1.3 nm thick CoFeB as its origin. Because of abrupt magnetic switching, we limit our discussion to effective magnetic fields in W/CoFeB bilayers with $t_\mathrm{W}$ = 2 nm.

\begin{figure}
\centering
\includegraphics[width=\columnwidth]{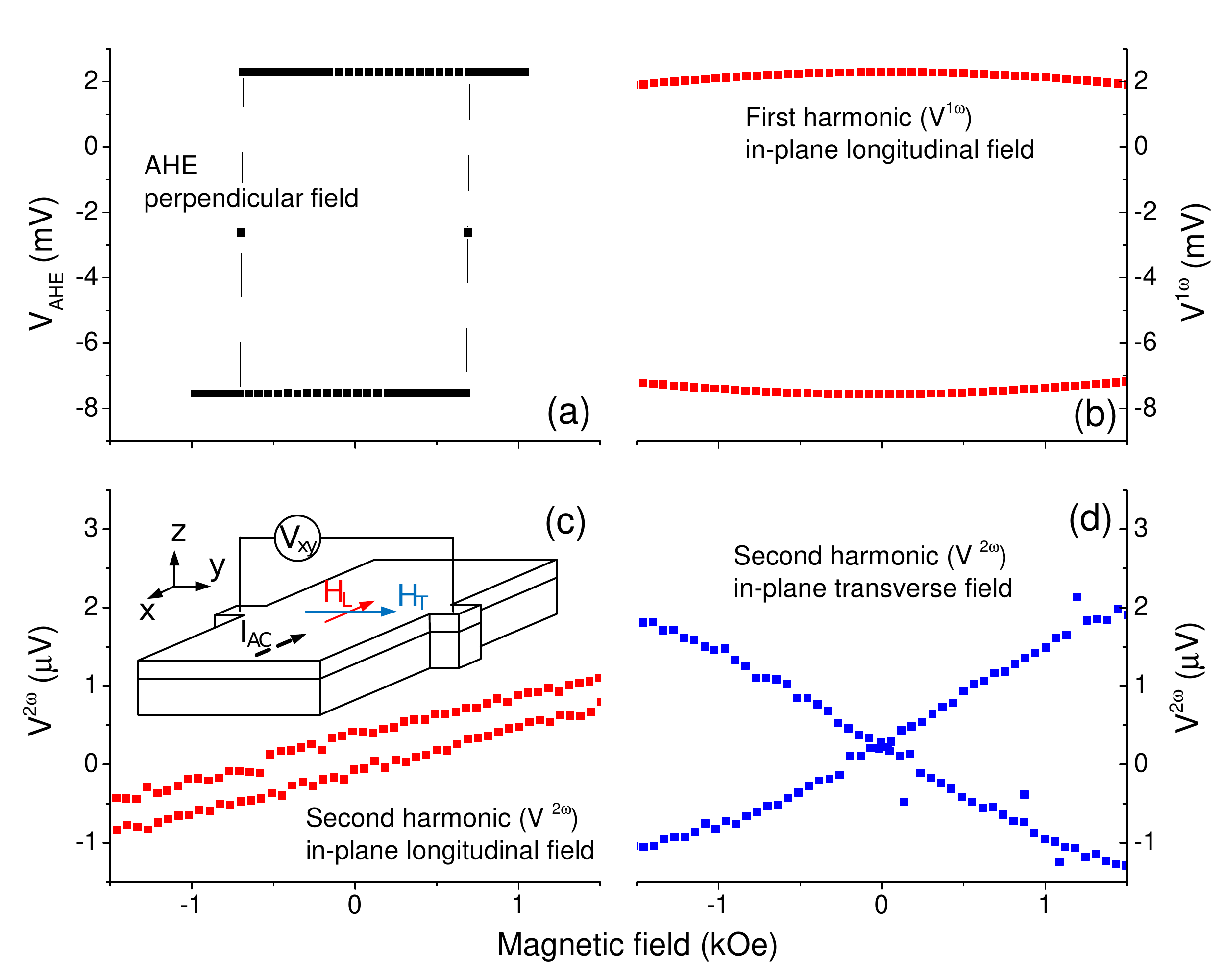}
\caption{(a) AHE vs. perpendicular magnetic field, (b) first harmonic Hall voltage vs. longitudinal in-plane field, (c) and (d) second harmonic Hall voltage vs. longitudinal and transverse field, respectively, obtained at T = 19 K for $t_\mathrm{W}$ = 2 nm. Note that the first Harmonic Hall voltage signal vs. transverse field is similar to (b). Inset in (c) shows the harmonic Hall voltage measurement configuration.}
\label{fig:hall}
\end{figure}

Figure \ref{fig:hall} shows AHE and harmonic Hall voltages at T = 19 K for the sample with $t_\mathrm{W}$ = 2 nm. The first harmonic Hall voltage signal measured in a magnetic field applied longitudinal to the Hall bar ($H_\mathrm{L}$) exhibits a parabolic shape and was fitted using a quadratic function both for the remanent magnetization point along +$z$ and -$z$ directions. Results for a magnetic field applied transverse to the long axis of the hall bar ($H_\mathrm{T}$) are almost identical. Likewise, the second harmonic signal was fitted using a linear function for both remanent magnetic field orientations. In this case, the two fitted lines are either symmetric (measurement along $H_\mathrm{L}$ - Fig. \ref{fig:hall}(c)) or asymmetric (measurement along $H_\mathrm{T}$ - Fig. \ref{fig:hall}(d)) with respect to the magnetic field polarity. The model presented in Ref. \cite{kim_spin-torque_2012} was used to calculate the effective longitudinal $\Delta H_\mathrm{L}$ and transverse $\Delta H_\mathrm{T}$ magnetic fields:

\begin{equation}
\Delta H_\mathrm{L(T)}=-2{\dfrac{\partial V^{2\omega}}{\partial H_\mathrm{L(T)}}}/{\dfrac{\partial^2 V^{1\omega}}{\partial H_\mathrm{L(T)}^2}}
\label{eq:field}
\end{equation}

\begin{figure}
\centering
\includegraphics[width=\columnwidth]{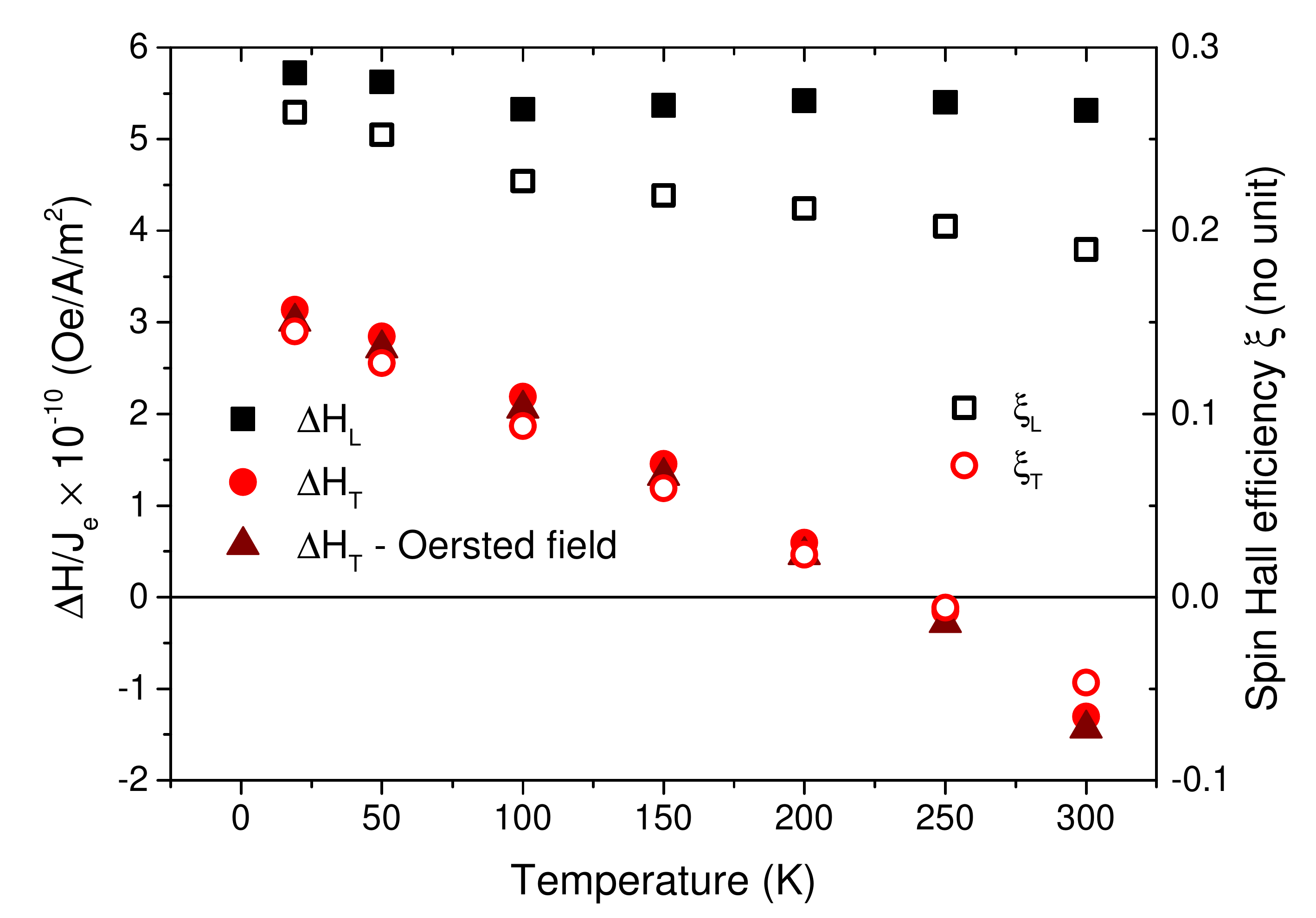}
\caption{Effective longitudinal $\Delta H_\mathrm{L}$ and transverse $\Delta H_\mathrm{T}$ fields as a function of temperature - left axis, measured for a 10-$/mum$-wide stripe with $t_\mathrm{W}$ = 2 nm. The Oersted field contribution was calculated from the current flowing through the buffer and subtracted from the transverse field. The right axis presents data of spin Hall efficiency calculated using Eq. \ref{eq:efficiency}.}
\label{fig:torque}
\end{figure}

Using a parallel resistor model, we calculated the current density in the W buffer layer as $J_\mathrm{c}$ = 2.17$\times$10$^{10}$ A/m$^2$. $J_\mathrm{c}$ drops by about 3\% at T = 19 K with respect to the room temperature value, as the resistance increases with decreasing temperature. This negative temperature coefficient is explained by the existence of an amorphous phase in the W buffer and will be discussed in detailed elsewhere. The data presented here were measured in 10-$\mu$m-wide stripe, however, no significant dependence of the strip's width on the determined values of effective magnetic field were found. Instead of $\theta_\mathrm{H}$, we characterized our bilayers using damping-like and field-like spin-orbit torque efficiencies ($\xi_\mathrm{L}$ and $\xi_\mathrm{L}$, respectively), as we independently measured these two torque components. 
The following equation was used to calculate the effective spin-orbit torques:

\begin{equation}
\Delta H_\mathrm{L(T)}/J_\mathrm{e}^W = \hbar \xi_\mathrm{L(T)}/2e M_\mathrm{s} t'_\mathrm{CoFeB}
\label{eq:efficiency}
\end{equation}

Where $\hbar$ is the reduced Planck's constant, 
$e$ is the electron charge, $M_\mathrm{s}$ is the saturation magnetization and $t'_\mathrm{CoFeB}$ is the effective thickness of CoFeB: $t'_\mathrm{CoFeB}$ = 0.92 nm (comparing to $t_\mathrm{CoFeB}$ = 1.3 nm nominal thickness, without a magnetic dead layer taken into account) \cite{skowronski_underlayer_2015}. The saturation magnetization of CoFeB at room temperature equals $\mu_0M_\mathrm{s}$ = 1.6 T and it increases to 2 T at T = 19 K. Figure \ref{fig:torque} presents the effective magnetic fields and spin-orbit torque efficiencies vs. temperature. 

We note that a relatively large planar Hall effect (PHE) of maximum 30\% of AHE value \cite{cho_large_PHE_2015} was measured in this sample. However, its magnitude vanishes when the magnetic field is aligned along the $x$ or $y$ directions, and therefore, we did not take it into account in the analysis of the effective magnetic fields. 

\begin{figure}
\centering
\includegraphics[width=\columnwidth]{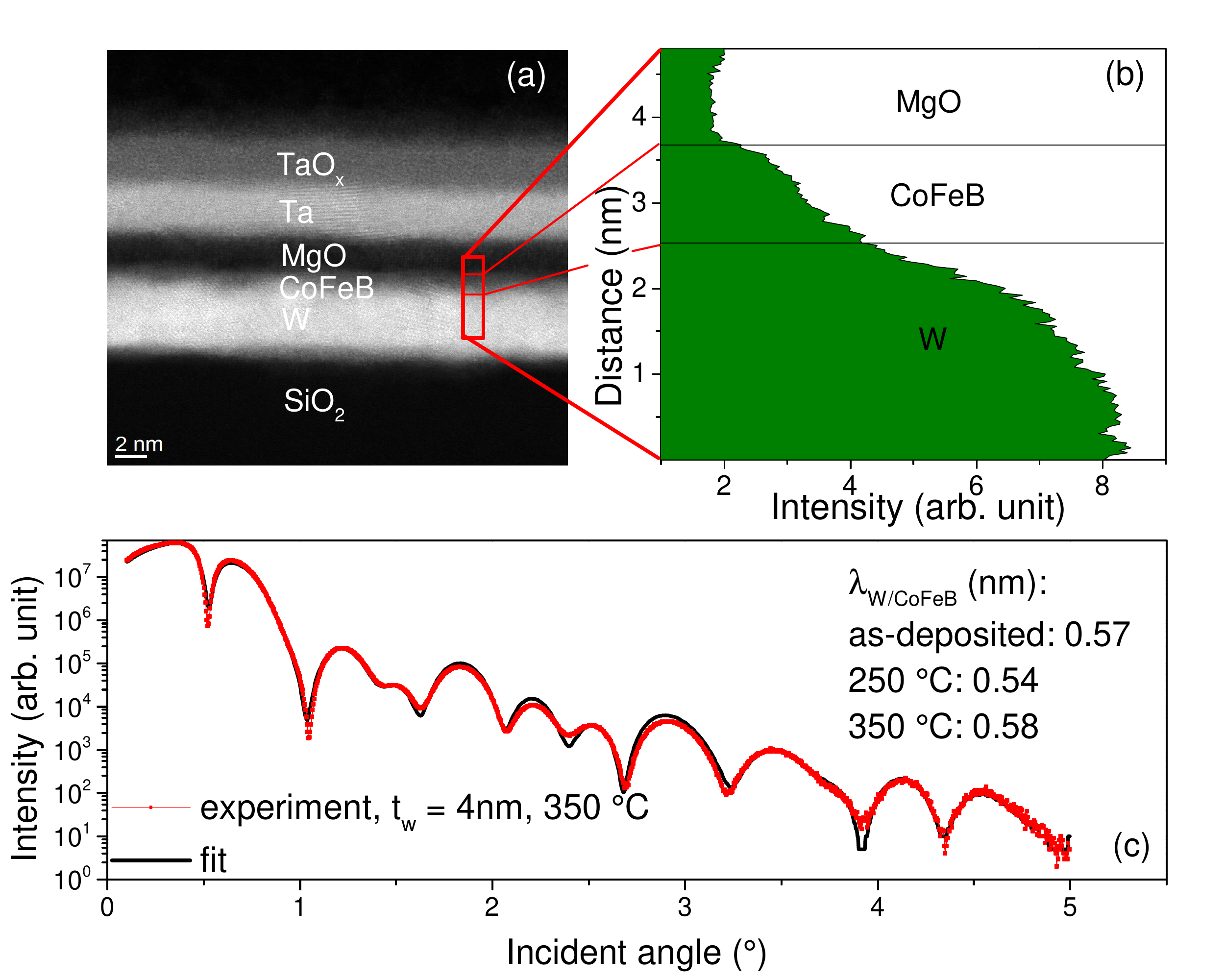}
\caption{Z-contrast image of fabricated multilayer stack (a) and line scan profile along vertical axis (b), indicating chemical intermixing at the interface between CoFeB and W. 
Good fits to XRR measurement are only obtained if large W/CoFeB interface roughness $\lambda_\mathrm{W/CoFeB}$ is assumed (c). The interface roughness does not change during annealing.}
\label{fig:tem}
\end{figure}

The result of $\xi_\mathrm{L}$ are in agreement with spin Hall angles derived in the literature data for W buffer layers \cite{cho_large_2015, pai_enhancement_2014, hao_giant_2015}. A slightly higher value is expected for thicker W, as $t_\mathrm{W}$ = 2 nm used here is smaller than the spin diffusion length of W \cite{hao_giant_2015}. The temperature dependence of the longitudinal spin-orbit-torque is similar to Ta \cite{kim_anomalous_2014}. On the other hand, the transverse effective field changes sign at around 250 K, which is not the case for Ta with the same thickness. Similar behavior has only been measured for very thin Ta, namely $t_\mathrm{Ta}$ = 0.5 nm, which indicates a much stronger interface effect in W. 
Generally, interface interactions (for example the Rashba effect) are dominant for thin buffers, whereas, for thicker heavy metals stronger contributions from field-like spin torques originating in the bulk are expected. Therefore, we conclude that in our case ($t_\mathrm{W}$ = 2 nm), both effects are of similar magnitude. This also implies, that the field-like spin-orbit torque opposes the interfacial interactions and that it varies more strongly with temperature. 

To elucidate the origin of the strong interface effect, TEM analysis of the samples was performed. Figure \ref{fig:tem} presents a Z-contrast image. The chemical sensitivity of the high-angle scattered electron signal provides good contrast between the polycrystalline W buffer, CoFeB/MgO bilayer and the Ta film with a thin oxidized layer on top. 
The line scan in Fig. \ref{fig:tem}(b) shows a gradual change of Z-contrast near the W/CoFeB interface, providing proof of considerable intermixing between these two layers. This observation is further corroborated by the XRR measurements of Fig. \ref{fig:tem}(c), which show that the W/CoFeB interface is very rough ($\lambda_\mathrm{W/CoFeB}$ = 0.57 nm), whereas a much smaller roughness is obtained for the CoFeB/MgO interface ($\lambda_\mathrm{CoFeB/MgO}$ = 0.18 nm). We note that $\lambda_\mathrm{W/CoFeB}$ is not affected by thermal annealing of the sample, i.e., intermixing primarily occurs during film growth.

Strong intermixing between the W and CoFeB layers, explains the large interface contribution (for example via the Rashba effect), to the total spin-orbit effective magnetic field in this study. Moreover, it can also explain high spin-orbit interactions in W/CoFeB bilayer leading to strong planar and spin Hall effects as well as anisotropic and spin-Hall magnetoresistance \cite{cho_large_2015}.

In summary, we investigated spin-orbit torques in perpendicularly magnetized CoFeB on thin W buffer layers. Harmonic Hall voltage measurements were used to determine longitudinal and transverse spin-orbit effective magnetic fields. The damping-like spin-orbit torque component is found to slightly increase with decreasing temperature reaching $\xi_\mathrm{L}$ = 0.27 at 19 K. In contrast, the field-like component, depends strongly on temperature and changes sign at 250 K. This temperature dependence indicates a strong interface contribution to the total spin-orbit torque (Rashba effect). From TEM and XRR measurements, we conclude that the large interface effect originates from strong intermixing between the W and CoFeB layers.  


\section*{Acknowledgments}
We thank J. Barnaś and T. Taniguchi for a fruitful discussion and J. Chęciński for help in calculations. This work is partially supported by the National Science Center, Poland, grant Harmonia No. UMO-2012/04/M/ST7/00799. W.S. acknowledges the National Science Center, Poland, grant No. UNO-2015/17/D/ST3/00500. S.v.D. acknowledges financial support from the European Research Council (ERC-2012-StG 307502-E-CONTROL). L.Y. acknowledges financial support from the Academy of Finland (Grant Nos. 286361 and 293929). Microfabrication was performed at Academic Center for Materials and Nanotechnology of AGH University. TEM analysis was conducted at the Aalto University Nanomicroscopy Center (Aalto-NMC).

\bibliographystyle{apl}

\end{document}